# Method of Estimation in the Presence of Non-response and Measurement Errors Simultaneously

Prayas Sharma and Rajesh Singh

Department of Statistics, Banaras Hindu University

Varanasi-221005, India

**Abstract**

The present paper discusses the problem of estimating the finite population mean of study variable in simple random sampling in the presence of non-response and response error together. The estimators in this article use auxiliary information to improve efficiency and we suppose that non–response and measurement error are present in both the study and auxiliary variables. A class of estimators has been proposed and its properties are studied in the simultaneous presence of non-response and response errors. It has been shown that proposed class of estimators is more efficient than the usual unbiased estimator, ratio and product estimators under non-response and response error together. In addition, a numerical study is carried out to compare the performance of the proposed class of estimators over others

## 1. Introduction

Over the past several decades, statisticians are peeping towards the problem of estimation of parameters in the presence of response error (measurement errors). In survey sampling, the properties of the estimators based on data usually presuppose that the observations are the correct measurements on characteristics being studied. However, this assumption is not satisfied in many applications and data is contaminated with measurement errors, such as reporting errors and computing errors. These measurement errors make the result invalid, which are meant for no measurement error case. If measurement errors are very small and we can neglect it, then the statistical inferences based on observed data continue to remain valid. On the contrary when they are not appreciably small and negligible, the inferences may not be simply invalid and inaccurate but may often lead to unexpected, undesirable and unfortunate consequences (see Srivastava and Shalabh (2001)). Some important sources of measurement errors in survey data are discussed in Cochran (1968). Shalabh (1997), Sud and Srivastva (2000). Singh and Karpe ( 2008, 2010), Kumar et al. (2011), Sharma and Singh (2013) studied the properties of some estimators of population mean under measurement error.

Consider a finite population U= ($U_1$, $U_2$, ........ $U_N$) of N units. Let Y and X be the study variate and auxiliary variate, respectively. Suppose that we have a set of n paired observations obtained through simple random sampling procedure on two characteristics X and Y. Further it is assumed that $x_i$ and $y_i$ for the $i^{th}$ sampling units are observed with measurement error instead of their true values ($X_i$, $Y_i$) For a simple random sampling scheme, let ($x_i$, $y_i$) be observed values instead of the true values ($X_i$, $Y_i$) for $i^{th}$ (i=1.2….n) unit, as

$$u_i = y_i - Y_i \quad (1.1)$$

$$v_i = x_i - X_i \quad (1.2)$$

Where $u_i$ and $v_i$ are associated measurement errors which are stochastic in nature with mean zero and variances $\sigma_u^2$ and $\sigma_v^2$, respectively. Further, let the $u_i$'s and $v_i$'s are uncorrelated although $X_i$'s and $Y_i$'s are correlated.

Let the population means of X and Y characteristics be $\mu_x$ and $\mu_y$, population variances of (x, y) be ($\sigma_X^2$, $\sigma_y^2$) and let $\rho$ be the population correlation coefficient between x and y respectively (see Manisha and Singh (2002)).

In sample surveys, the problem of non-response is common and is more widespread in mail surveys than in personal interviews. The usual approach to overcome non-response problem is to contact the non-respondent and obtain the information as much as possible. Hansen and Hurwitz (1946) were the first to deal with the problem of non-response they proposed a sampling scheme that involves taking a subsample of non responds after the first mail attempt and the obtain the information by personal interview.

For a finite population $U = \{U_1, U_2....U_N\}$ of size N and a random sample of size n is drawn without replacement. Let the characteristics under study, say, y takes value $y_i$ on the unit $U_i (i = 1,2,...N)$. In survey on human population it is often the case that $n_1$ unit respond on the first attempt while $n_1$ (=n- $n_1$) units do not provide any response. In the case of non-response of at initial stage Hansen and Hurwitz (1946) suggested a double sampling plan for estimating the population mean comprising the following steps :

(i) A simple random sample of size n is drawn and the questionnaire is mailed to the sample units;

(ii) A sub-sample of size r = $(n_2/k)$,(k>1) from the $n_2$ non responding units in the initial step attempt is contacted through personal interviews.

Note that Hansen and Hurwitz (1946) considered the mail surveys at the first attempt and the personal interviews at the second attempt. In the Hansen and Hurwitz method the population is supposed to be consisting of response Stratum of size $N_1$ and the non response stratum of size $N_2$ =(N-$N_1$). Let $\overline{Y} = \sum_{i=1}^{N} y_i /N$ and $S_y^2 = \sum_{i=1}^{N}(y_i - \overline{Y})^2/(N-1)$ denote the mean and the population variance of the study variable y. Let $\overline{Y}_1 = \sum_{i=1}^{N_1} y_i /N_1$ and $S_{y(1)}^2 = \sum_{i=1}^{N_1}(y_i - \overline{Y})^2/(N_1-1)$ denote the mean and variance of response group. Similarly, let $\overline{Y}_2 = \sum_{i=1}^{N_2} y_i /N_2$ and $S_{y(2)}^2 = \sum_{i=1}^{N_2}(y_i - \overline{Y})^2/(N_2-1)$ denote the mean and variance of the non-response group. The population mean can be written as $\overline{Y} = W_1\overline{Y}_1 + W_2\overline{Y}_2$, where $W_1$ = ($N_1$/N) and $W_2$ = ($N_2$/N). The sample mean $\overline{y}_1 = \sum_{i=1}^{n_1} y_i /n_1$ is an unbiased for $\overline{Y}_1$, but has a bias equal to $W_2(\overline{Y}_1 - \overline{Y}_2)$ in estimating the population mean $\overline{Y}$.

The sample mean $\overline{y}_{2r} = \sum_{i=1}^{r} y_i /r$ is unbiased for the mean $y_2$ for the $n_2$ units. Hansen and Hurwitz (1946) suggested an unbiased estimator for the population mean $\overline{Y}$ is given by

$$\overline{y}^* = w_1\overline{y}_1 + w_2\overline{y}_{2r}$$

Where $w_1$ = ($n_1$/n) and $w_2$ = ($n_2$/n) are responding and non-responding proportions in the sample. The variance of $\overline{y}^*$ is given by

$$V(\overline{y}^*) = \left(\frac{1-f}{n}\right)S_y^2 + \frac{W_2(k-1)}{n}S_{y(2)}^2 \text{ ; where f=(n/N).}$$

In the sampling literature, it is well known that efficiency of the estimator of population mean of a study variable y can be increased by the use of auxiliary information related to x which is highly correlated with study variable y. Cochran (1977) suggested the ratio and regression estimator of the population mean $\overline{Y}$ of study variable y in which information on the auxiliary variable is obtained from all sample units, and the population mean of auxiliary variable x is known, while some units do not provide any information on study variable y . Rao(1986), Khare and Srivastava (1995,1997), Okafor and Lee (2000)  and Singh and Kumar (2008,2009,2010) have suggested some estimator for population mean of the study variable y using auxiliary information in presence of non response.

Let $x_i, (i = 1, 2....N)$ denote a auxiliary characteristics correlated with the study variable $y_i, (i = 1, 2....N)$ the population mean of auxiliary variable is $\overline{X} = \sum_{i=1}^{N} x_i / N$. Let $\overline{X}_1$ and $\overline{X}_2$ denote the population means of the response and non-response groups. Let $\overline{x}_1 = \sum_{i=1}^{n_2} x_i / n_2$

$\overline{x}_2 = \sum_{i=1}^{n_2} x_i / n_2$, $\overline{x}_{2r} = \sum_{i=1}^{r} x_i / r$ denote the means of the $n_1$ responding units , $n_2$ non-responding units, and $r=(n_2/k)$ sub-sampled units respectively. In this paper we have merged two major concepts for improvement of estimation techniques that is consideration of measurement error and non-response in the estimation procedure and proposed a class of estimators.

**Notations**

Let $\overline{x} = \frac{1}{n} \sum_{i=1}^{n} x_i$, $\overline{y} = \frac{1}{n} \sum_{i=1}^{n} y_i$, be the unbiased estimator of population means $\overline{X}$ and $\overline{Y}$, respectively but $s_x^2 = \frac{1}{n-1} \sum_{i=1}^{n} (x_i - \overline{x})^2$ and $s_y^2 = \frac{1}{n-1} \sum_{i=1}^{n} (y_i - \overline{y})^2$ are not unbiased estimator of ($\sigma_x^2$, $\sigma_y^2$), respectively. The expected values of $s_x^2$ and $s_y^2$ in the presence of measurement error are, given by,

$E(s_x^2) = \sigma_x^2 + \sigma_v^2$

$E(s_y^2) = \sigma_y^2 + \sigma_u^2$

And for non-response group

$$E(s_{x_2}^2) = \sigma_{x_2}^2 + \sigma_{v_2}^2$$

$$E(s_{y_2}^2) = \sigma_{y_2}^2 + \sigma_{u_2}^2$$

When the error variance $\sigma_v^2$ is known, the unbiased estimator of $\sigma_x^2$, is $\hat{\sigma}_x^2 = s_x^2 - \sigma_v^2 > 0$, and when $\sigma_u^2$ is known, then the unbiased estimator of $\sigma_y^2$ is $\hat{\sigma}_y^2 = s_y^2 - \sigma_u^2 > 0$.

Similarly, for the non response group the unbiased estimator of $\sigma_{x_2}^2$, is $\hat{\sigma}_{x_2}^2 = s_{x_2}^2 - \sigma_{v_2}^2 > 0$, and when $\sigma_{u_2}^2$ is known, then the unbiased estimator of $\sigma_{y_2}^2$ is $\hat{\sigma}_{y_2}^2 = s_{y_2}^2 - \sigma_{u_2}^2 > 0$.

$$E(s_{x_2}^2) = \sigma_{x_2}^2 + \sigma_{v_2}^2$$

$$E(s_{y_2}^2) = \sigma_{y_2}^2 + \sigma_{u_2}^2$$

We define

$$\bar{y} = \mu_y(1 + e_0)$$

$$\bar{x} = \mu_x(1 + e_1)$$

Such that

$$E(e_0) = E(e_1) = 0,$$

And up to the first degree of approximation (when finite population correction factor is ignored)

$$E(e_0^2) = \frac{C_y^2}{n}\left(1 + \frac{S_u^2}{S_y^2}\right) + \frac{W_2(k-1)}{n}C_{y_2}^2\left(1 + \frac{S_{u_2}^2}{S_{y_2}^2}\right)$$

$$E(e_1^2) = \frac{C_x^2}{n}\left(1 + \frac{S_v^2}{S_x^2}\right) + \frac{W_2(k-1)}{n}C_{x_2}^2\left(1 + \frac{S_{v_2}^2}{S_{x_2}^2}\right)$$

$$E(e_0 e_1) = \frac{\rho_{yx} C_y C_x}{n} + \frac{W_2(k-1)}{n}\rho_{yx_2} C_{y_2} C_{x_2}$$

$$C_y = S_y/\bar{Y}, \quad C_x = S_x/\bar{X}, \quad C_{y_2} = S_{y_2}/\bar{Y}, \quad C_{x_2} = S_{x_2}/\bar{X}$$

$$\rho_{xy} = S_{xy}/S_x S_y$$

## 2. Adapted estimator

A traditional estimator for estimating population mean in the simultaneous presence of response and non-response error is given by,

$$t_1 = \bar{y} \tag{2.1}$$

Expression (2.1) can be written as

$$t_1 - \bar{Y} = \bar{Y}^2(1 + e_0) \tag{2.2}$$

Taking expectation both sides of (2.2), we get bias of estimator $t_1$ given as

$$\text{Bias}(t_1) = 0 \tag{2.3}$$

Squaring both sides of (2.2) we have

$$(t_1 - \bar{Y})^2 = \bar{Y}^2 e_0^2$$

Taking expectation and using notations, we get the mean square error of $t_1$ up to first order of approximation, as

$$\text{MSE}(t_1) = \frac{S_y^2}{n}\left(1 + \frac{\sigma_u^2}{S_y^2}\right) + AS_{y2}^2\left(1 + \frac{\sigma_{u2}^2}{S_{y2}^2}\right) \tag{2.4}$$

Or

$$\text{MSE}(t_1) = M \tag{2.5}$$

Where, $A = \dfrac{(k-1)W_2}{n}$ and $M = \dfrac{S_y^2}{n}\left(1 + \dfrac{\sigma_u^2}{S_y^2}\right) + AS_{y2}^2\left(1 + \dfrac{\sigma_{u2}^2}{S_{y2}^2}\right)$

In the case, when the measurement error is zero or negligible, MSE of estimator $t_1$ is given by,

$$\text{MSE}^*(t_1) = \frac{S_y^2}{n} + AS_{y2}^2 \tag{2.6}$$

where, $M_{t_1} = \dfrac{\sigma_u^2}{n} + A\sigma_{u2}^2$ is the contribution of measurement errors in $t_1$.

When there is non-response and response error both are present, a ratio type estimator for estimating population mean is, given by

$$t_r = \frac{\bar{y}^*}{\bar{x}^*}\bar{X} \tag{2.7}$$

Expressing the estimator $t_r$ in terms of e's we have

$$t_r = \bar{Y}(1+e_0)(1+e_1)^{-1} \tag{2.8}$$

Expanding equation (2.8) and simplifying, we have

$$(t_r - \bar{Y}) = \bar{Y}[e_0 - e_1 - e_0e_1 + e_1^2] \tag{2.9}$$

Taking expectation both sides of (2.9) we get the bias of estimator $t_r$ given as,

$$\text{Bias}(t_r) = \left[\frac{S_x^2}{n}\left(1+\frac{\sigma_v^2}{S_x^2}\right) + AS_{x2}^2\left(1+\frac{\sigma_{v2}^2}{S_{x2}^2}\right) - 2\left(\frac{1}{n}\rho_{xy}S_xS_y + A\rho_{xy2}S_{x2}S_{y2}\right)\right] \tag{2.10}$$

Squaring both sides of (2.9), we have

$$(t_r - \sigma_y^2)^2 = \bar{Y}^2[e_0^2 + e_1^2 - 2e_0e_1] \tag{2.11}$$

Taking expectations of (2.11) and using notations, we get the MSE of estimator $t_r$ as

$$\text{MSE}(t_r) = \frac{1}{n}\left[S_y^2\left(1+\frac{\sigma_u^2}{S_y^2}\right) + S_x^2\left(1+\frac{\sigma_v^2}{S_x^2}\right) - 2\rho_{xy}S_xS_y\right]$$

$$+ A\left[S_{y2}^2\left(1+\frac{\sigma_{u2}^2}{S_{y2}^2}\right) + S_{x2}^2\left(1+\frac{\sigma_{v2}^2}{S_{x2}^2}\right) - 2\rho_{xy2}S_{x2}S_{y2}\right] \tag{2.12}$$

$$= \left[\frac{S_y^2}{n}\left(1+\frac{\sigma_u^2}{S_y^2}\right) + AS_{y2}^2\left(1+\frac{\sigma_{u2}^2}{S_{y2}^2}\right) + \frac{S_x^2}{n}\left(1+\frac{\sigma_v^2}{S_x^2}\right) + AS_{x2}^2\left(1+\frac{\sigma_{v2}^2}{S_{x2}^2}\right) - 2\left(\frac{1}{n}\rho_{xy}S_xS_y + A\rho_{xy2}S_{x2}S_{y2}\right)\right]$$

$$= M+N-2O \tag{2.13}$$

Where, $M = \left\{\frac{1}{n}S_y^2\left(1+\frac{\sigma_u^2}{S_y^2}\right) + AS_{y2}^2\left(1+\frac{\sigma_{u2}^2}{S_{y2}^2}\right)\right\}$

$$N = \left\{\frac{1}{n}S_x^2\left(1+\frac{\sigma_v^2}{S_x^2}\right) + AS_{x2}^2\left(1+\frac{\sigma_{v2}^2}{S_{x2}^2}\right)\right\}$$

$$O = \left\{\frac{1}{n}\rho_{xy}S_xS_y + A\rho_{xy2}S_{x2}S_{y2}\right\}$$

A regression estimator under measurement error and non-response is defined as

$$t_{lr} = \bar{y}^* + b(\bar{X} - \bar{x}^*) \tag{2.14}$$

Expressing the estimator $t_r$ in terms of e's we have

$$t_{lr} = \bar{Y}(1+e_0) - b\bar{X}e_1$$

Expanding equation (2.14) and simplifying, we have

$$(t_{lr} - \bar{Y}) = (\bar{Y}e_0 - b\bar{X}e_1) \tag{2.15}$$

Squaring both sides of (2.15) and after simplification, we have

$$(t_{lr} - \bar{Y})^2 = \left[\bar{Y}^2 e_0^2 + b^2 \bar{X}^2 e_1^2 - 2b\bar{X}\bar{Y}e_0 e_1\right] \tag{2.16}$$

Taking expectations both sides of (2.16) we get the MSE of estimator $t_{lr}$ as

$$MSE(t_{lr}) = M + b^2 R^2 N - 2bRO \tag{2.17}$$

The optimum value of b is obtained by minimizing (2.17) and is given by

$$b^* = \frac{1}{R}\left[\frac{O}{N}\right] \tag{2.18}$$

Substituting the optimal value of b in equation (2.17) we obtain the minimum MSE of the estimator $t_{lr}$ as

$$MSE(t_{lr})_{min} = M\left[1 - \frac{O^2}{MN}\right] \tag{2.19}$$

In the case, when the measurement error is zero or negligible, MSE of estimator $t_1$ is given by,

$$MSE(t_{lr}) = \frac{1}{n}S_y^2\left[1-\rho_{xy}^2\right] + \frac{(k-1)W_2}{n}\left[S_{y2}^2 + b^2 S_{x2}^2 - 2b\rho_{xy2}S_{x2}S_{y2}\right] \tag{2.20}$$

**3. Proposed class of Estimator**

We propose a class of estimators given by

$$t_p = m_1 \bar{y}^* + m_2 \frac{\bar{y}^*}{\bar{x}^*} \bar{X} \qquad (3.1)$$

We note that

(1) For $(m_1, m_2) = (1,0)$ $t_1 = \bar{y}^*$ (usual unbiased estimator)

(2) For $(m_1, m_2) = (0,1)$ $t_2 = \frac{\bar{y}^*}{\bar{x}^*} \bar{X}$ (usual ratio estimator)

Thus the proposed class of estimators is generalised version of usual unbiased estimator and ratio estimator. Expressing the estimator $t_p$ in terms of e's we have

$$t_p = m_1 \bar{Y}(1+e_0) + m_2 \bar{Y}(1+e_0)(1+e_1)^{-1} \qquad (3.2)$$

Expanding equation (3.2) and simplifying, we have

$$(t_p - \bar{Y}) = \bar{Y}[e_0 + m_2(-e_1 + e_1^2 - e_0 e_1)] \qquad (3.3)$$

Squaring both sides of (3.3) and after simplification, we have

$$(t_p - \bar{Y})^2 = \bar{Y}^2 [e_0^2 + m_2^2 e_1^2 - 2m_2 e_0 e_1] \qquad (3.4)$$

Taking expectations of (3.4) and using notations, we get the MSE of estimator $t_r$ as

$$MSE(t_p) = M + m_2 R^2 N - 2m_2 RO \qquad (3.5)$$

The optimum value of $m_2$ is obtained by minimizing (3.5), given by

$$m_2^* = \frac{1}{R}\left[\frac{O}{N}\right] \qquad (3.6)$$

And $m_1^* = 1 - m_2^*$

Substituting the optimal value of $m_2$ in equation (3.5) we obtain the minimum MSE of the estimator $t_p$ as

$$MSE(t_p)_{min} = M\left[1 - \frac{O^2}{MN}\right] \qquad (3.7)$$

Minimum MSE of proposed class of estimator $t_p$ given in (3.7) is same as the MSE of regression estimator under simultaneous presence of non-response and measurement error, given in equation (2.19).

## 2. Efficiency Comparisons

First we compare the efficiency of the proposed estimator $t_p$ with usual unbiased estimator

$$MSE(t_1) - MSE(t_P)_{min} > 0$$

If

$$\left[ M - M\left(1 - \frac{O^2}{MN}\right) \right] > 0$$

$$\left[ \frac{O^2}{MN} \right] > 0 \qquad (4.1)$$

The condition listed in (4.1) shows that proposed family of estimators is always better than the usual estimator under the non-response and measurement error.

Next, we compare the ratio estimator with proposed family of estimators $t_p$,

$$MSE(t_2)_{min} - MSE(t_P)_{min} > 0$$

$$\left[ (M + N - 2O) - M\left(1 - \frac{O^2}{MN}\right) \right] > 0$$

$$[N - O]^2 > 0 \qquad (4.2)$$

We observe that the condition (4.2) holds always true and shows proposed family of estimators is always better than the Ratio estimator under the non-response and measurement error.

## 5. Empirical Study

**Data statistics:** The data used for empirical study was taken from Gujrati and Sangeetha (2007) -pg, 539. where,

$Y_i$ = True consumption expenditure,

$X_i$ = True income,

$y_i$ = Measured consumption expenditure,

$x_i$ = Measured income.

From the data given we get the following parameter values:

**Table.5.1: Value of the Parameters**

| n | $\mu_y$ | $\mu_x$ | $S_y$ | $S_x$ | $\rho$ | $\sigma_u^2$ | $\sigma_v^2$ |
|---|---------|---------|-------|-------|--------|--------------|--------------|
| 70 | 981.29 | 1755.53 | 613.66 | 1406.13 | 0.778 | 36.00 | 36.00 |
| $\mu_{y2}$ | $\mu_{x2}$ | $S_{y2}$ | $S_{x2}$ | $\rho_2$ | R | $W_2$ | |
| 597.29 | 1100.24 | 244.11 | 631.51 | 0.445 | 0.5589 | 0.25 | |

**Table (5.2): Showing the MSE of the estimators with and without measurement errors**

| Estimators | MSE Without Error | Contribution of meas. error in MSE | Contribution of non-response | MSE including me. Errors & non-response |
|---|---|---|---|---|
| $t_1 = \bar{y}^*$ | 10759.39 | 1.03 | 2553.840 | 13313.58 |
| $t_r$ | 6967.135 | 1.35 | 4607.335 | 11574.92 |
| $t_{lr}$ | 4246.903 | 0.86 | 2527.751 | 6775.036 |
| $t_p$ | 4246.903 | 0.86 | 2527.751 | 6775.036 |

Table (5.2) exhibits that measurement error and non-response plays an important role in increasing the MSE of an estimator. We also conclude that contribution of measurement error and non-response in usual estimator is less than in comparison to the ratio estimator; these observations have interesting implication where the ratio estimator performs better than sample mean under the absence of any measurement error in X characteristics. There may be a case when ratio estimator is poor than sample mean under the consideration of any measurement error. It is observed from Table (5.2) that the performance of our proposed estimator $t_p$ is better than usual estimator $t_1$ and ratio estimator $t_r$ under non-response and measurement error. Further it is observed that contribution of non-response error is larger than the response error in increasing the MSE of the estimators.

**Conclusion**

In this present study we have suggested a class of estimator of the population mean of study variable y using auxiliary information. The estimators in this article use auxiliary information to improve efficiencies and we suppose that non–response and measurement error are present in both the study and auxiliary variables. In addition, some known estimator of population mean such as usual unbiased estimator and ratio estimator for population mean are found to be members of the proposed class of estimators .We have obtained the MSEs of the proposed class of estimators up to the first order of approximation in the simultaneous presence of non-response and response error. The proposed class of estimators are advantageous in the sense that the properties of the estimators which are members of the proposed class of estimators can be easily obtained from the properties of the proposed class of estimators .In theoretical and empirical comparisons we have shown that the proposed class of estimators are more efficient than the usual unbiased estimator and ratio estimator and equally efficient to regression estimator under non-response and measurement error together.